\documentclass{article}

\usepackage{arxiv}

\usepackage[utf8]{inputenc} % allow utf-8 input
\usepackage[T1]{fontenc}    % use 8-bit T1 fonts
\usepackage{hyperref}       % hyperlinks
\usepackage{url}            % simple URL typesetting
\usepackage{booktabs}       % professional-quality tables
\usepackage{amsfonts}       % blackboard math symbols
\usepackage{nicefrac}       % compact symbols for 1/2, etc.
\usepackage{microtype}      % microtypography
\usepackage{cleveref}       % smart cross-referencing
\usepackage{lipsum}         % Can be removed after putting your text content
\usepackage{graphicx}
\usepackage{natbib}
\usepackage{doi}

\title{Towards dialogue based, computer aided software requirements elicitation.}

\newif\ifuniqueAffiliation
\uniqueAffiliationtrue

\ifuniqueAffiliation % Standard variant of author block
\author{ \href{https://orcid.org/0000-0002-7121-6816}{\includegraphics[scale=0.06]{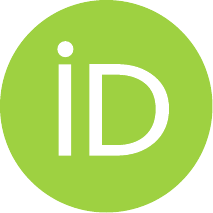}\hspace{1mm}Vasiliy Seibert}\\
	Institute for software and systems engineering\\
	TU-Clausthal\\
	38678, Clausthal-Zellerfeld \\
	\texttt{vasiliy.seibert@tu-clausthal.de} \\
}
\else
\usepackage{authblk}

\newbox{\orcid}\sbox{\orcid}{\includegraphics[scale=0.06]{orcid.pdf}} 
\author[1]{%
	\href{https://orcid.org/0000-0002-7121-6816}{\usebox{\orcid}\hspace{1mm}Vasiliy Seibert\thanks{\texttt{vasiliy.seibert@tu-clausthal.de}}}%
}
\fi

\renewcommand{\shorttitle}{Towards dialogue based, computer aided software requirements elicitation.}

\hypersetup{
pdftitle={Towards dialogue based, computer aided software requirements elicitation},
pdfsubject={q-bio.NC, q-bio.QM},
pdfauthor={Vasiliy Seibert},
pdfkeywords={Requirements Engineering, NLP, LLM},
}

\begin{document}
\maketitle

\begin{abstract}
	Several approaches have been presented, which aim
to extract models from natural language specifications.
These approaches have inherent weaknesses for they
assume an initial problem understanding that is
perfect, and they leave no room for feedback.
Motivated by real-world collaboration settings between
requirements engineers and customers, this paper
proposes an interaction blueprint that aims for dialogue
based, computer aided software requirements analysis.
Compared to mere model extraction approaches,
this interaction blueprint encourages individuality,
creativity and genuine compromise. A simplistic
Experiment was conducted to showcase the general
idea. This paper discusses the experiment as well
as the proposed interaction blueprint and argues,
that advancements in natural language processing and
generative AI might lead to significant progress in a
foreseeable future. However, for that, there is a need to move away from a magical black box expectation and
instead moving towards a dialogue based approach that
recognizes the individuality that is an undeniable part of
requirements engineering.
\end{abstract}

% keywords can be removed

\section{Introduction}

The transition from software requirements descriptions, written in natural language, to domain models requires skill and holds the danger of making costly mistakes. It is also a task that can take up much time and resources. Several approaches have been presented, which aim to automatically (or semi-automatically) extract domain models from natural language specifications (\cite{survey_1,survey_2}). The expected inputs for these approaches are highly heterogeneous and range from restricted/unrestricted natural language requirements descriptions to outright structured text elements. The outputs are likewise heterogeneous, including UML class, object, sequence and use-case diagrams. Proposed methods and tools for realizing this extraction, generally, try to identify relevant concepts and meaningful relationships between them and afterwards mapping the results to UML elements. For this, they utilize data driven as well as rule based approaches.

The authors \cite{libya} present an approach for extracting UML class diagrams from pre-edited natural language requirements. This approach requires the user/customer to simplify the input texts using a set of rules (e.g. rephrasing the input sentence to match a Subject-Predicate-Object sentence structure). Using these pre-edited sentences, the authors propose a second set of rules to extract relevant concepts as well as meaningful relationships between them. This approach utilizes the Stanford Core NLP API for analyzing and annotating the input texts. The results were evaluated by comparison with a similar, well cited (\cite{umgar}) approach.

The authors \cite{elallaoui2018automatic} present an approach for automatically transforming user stories to UML use case diagrams, using a combination of NLP methods, combined with an algorithmic approach. The authors motivate, that the majority of research focuses on extracting UML models from requirement documents and therefore agile development is underrepresented. 

The survey conducted by \cite{survey_1} analyzed 70 selected Papers that aim to automatically (or
semi-automatically) extract UML models from natural language text, with the objective, of better understanding the initial problem scenarios, used techniques, result evaluation and described challenges. The authors reflected their findings and derived a concept framework for further research. This proposed concept framework aims to unify further research by abstracting future tools/approaches/solutions (for the task of extracting UML models from natural language text) into 5 essential steps: \textbf{(i) natural language} every future work will start with a requirements description written in natural text. This document can be seen as the initial Dataset, \textbf{(ii) natural language processing} the initial Dataset is then processed using natural language processing. This step could include advancements in lemmatization, tokenization, utilizing various word-embeddings, named-entity-recognition, dependency parsing etc. The progress of extracting UML models from natural language text is therefore significantly linked to the progress made in natural language processing. \textbf{(iii) identification} the resulting sentence graph is then further analyzed, e.g. by utilizing heuristic rules. An example for a reoccurring heuristic rule can be found in \cite{libya}:
\hfill \break

\textit{"C-Rule 1: Extract the common nouns (...), and proper nouns (...) from the text and map them to classes."}
\hfill \break

The identification step aims to extract relevant concepts from the natural language text. In the \textbf{(iv) relation resolution} step the objective is to extract meaningful relations between the previously identified concepts. An example for a corresponding heuristic rule can be found in \cite{umgar}: \hfill \break

\textit{"The verb phrase between subject and object is taken as message passed between objects."}
\hfill \break

The last step of the concept framework \textbf{(v) unified modelling language} poses the desired output format.
\hfill \break

\begin{figure}
    \centering
	\includegraphics[clip,width=0.4\linewidth]{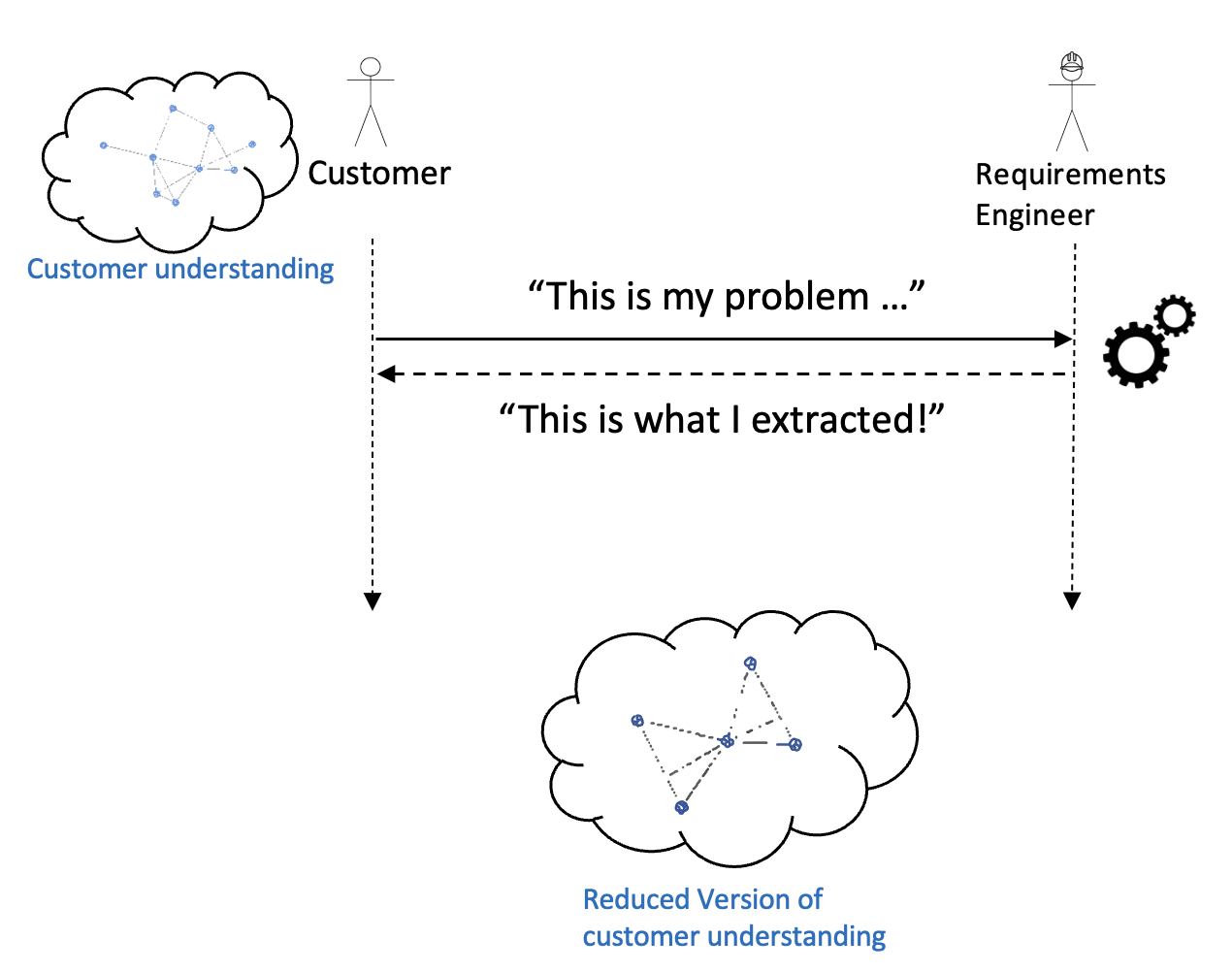}
	\caption{Interaction scheme for a model extraction approach.}
	\label{fig: Figure 1.}       % Give a unique label
\end{figure}

Although the motivation behind this branch of research is coherent, this paper argues that the current research that is conducted towards the extraction of models from natural language requirements also has fundamental weaknesses. (a) All the presented works as well as the proposed concept framework for further research \textbf{assumes} that the initial Dataset (requirements description in natural language) is complete and describes the underlying problem scenario \textbf{perfectly}. The quality of the requirements document is not being questioned by any of the presented works. Furthermore, (b) the extraction process offers no possibility for customer feedback. Essentially, the communication process between the customer and the extraction tool at hand can be depicted as in Figure 1, in which the customer starts the interaction process with an imagination of the problem scenario/ domain model, communicates it to the extraction tool (let's call it virtual requirements engineer) and eventually culminates with a reduced version of his preexisting understanding.

This communication process stands in contrast to how the interaction would look like in a professional, real world setting. In which a requirements engineer would listen to the customer's explanations and map them to his own experiences and knowledge. The requirements engineer would ask questions to verify his imagination of the problem scenario/ domain model and adjust it accordingly to the customer's response. The requirements engineer would propose new concepts and new relationships to which the customer could give his feedback to. He would try to enrich and abstract the contents of the discussion and formulate a domain model (This is what he was hired for). Furthermore, for the requirements engineer the interaction with a customer holds special value, since it is an opportunity for him to verify his domain knowledge. It is a chance for him to stand up to date with current advancements and needs of the customer. This new Information (or falsification of information, since it became irrelevant), can be carried on to the next customer. Eventually the customer and the requirements engineer would agree on a mutual understanding of the problem scenario/ domain model. This new model might differ from the understanding that the customer started the interaction with.

So the ultimate goal of the interaction process should be to generate a mutual understanding through dialogue rather than to extract a model from potentially incomplete descriptions. This paper reflects on current approaches towards the extraction of models from natural language requirements descriptions, it proposes a novel interaction blueprint which aims to generate an enriched model of the problem scenario/ domain model through dialogue, it illustrates the main idea with an experiment, followed by a discussion. 

\section{Mutual Problem understanding through Dialogue}
Inspired by realistic, professional collaboration settings between requirements engineers and customers, this paper proposes a novel interaction blueprint that aims for dialogue based, computer aided software requirements analysis. The objective is to generate a mutual understanding of the problem scenario/ domain model between a virtual requirements engineer and a given customer, rather than to extract such a model from potentially incomplete requirements descriptions in natural language. This interaction (see Figure 2) is initiated by the customer, who has an initial problem understanding. The customer communicates his problem scenario to a virtual requirements engineer, who then tries to understand the customer's explanations and tries to link them to his own knowledge base. The requirements engineer could acknowledge the customer's efforts by replying with reactions (e.g. "I know what you are talking about", "I don't know what you are talking about", "I know something similar to what you are talking about"). 

\begin{figure}[!htb]
    \centering
	\includegraphics[clip,width=0.4\linewidth]{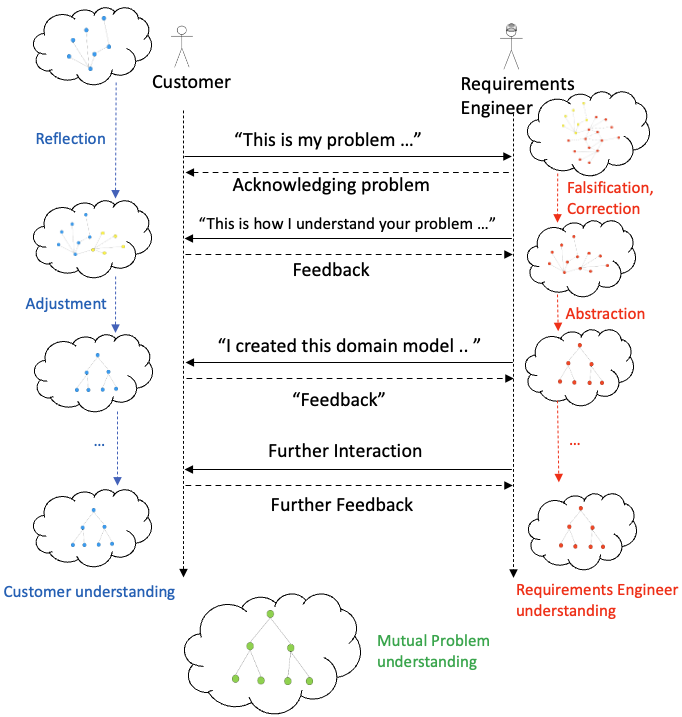}
	\caption{Proposed interaction blueprint.}
	\label{fig: Figure 2.}       % Give a unique label
\end{figure}

Another interaction step could be initiated by the virtual requirements engineer, by proposing the customer concepts and relationships based on the customer's explanations (e.g. "In addition to what you told me I would suggest ...", "This is how I understand your problem scenario ...", "I know about a domain model that has different concepts but similar structures ... "). This additional information can then be reviewed by the customer and based on his feedback (a) the virtual requirements engineer can alter/prioritize his knowledge base and (b) a mutual problem understanding emerges. This initiative might cause the customer to reflect on his previous understanding. For the requirements engineer, the corresponding Feedback might mean a correction, falsification of preexisting knowledge (something that worked for one customer must not work for the next one). By abstracting the contents of the discussion, the virtual requirements engineer introduces the concept of relevance. His preexisting knowledge (experience) enables him to generalize the discussed concepts and meaningfully relationships between them. Feedback from different customers, helps the virtual requirements engineer to acquire general knowledge, which might be his most valuable asset. The proposed domain model means for the customer, that he has to let go of details that he previously considered relevant. This adjustment must not mean a loss of value but rather a consolidation of information.

The described interaction steps can be repeated multiple times. Eventually the customer and the virtual requirements engineer would agree on a mutual problem understanding/ domain model. Furthermore, both the customer and the virtual requirements engineer would end (or halt) the interaction with a problem understanding/ knowledge base that differs from what they started the interaction with. \textbf{They both learned from each other during the course of the interaction.}

\section{Experiment}

For the purpose of demonstrating (a fraction of) the proposed interaction blueprint, an experiment was conducted (See Figure 3). First a corpus of four texts was collected from different sources. All of the texts describe the same topic (in this experiment the topic of choice was "Akita dog", however this experiment should work with any given topic). From the texts, all nouns were extracted. The extraction of nouns was performed by utilizing the Stanford Core NLP API, which offers POS (Part of Speech tagging) amongst other tools. The four lists of nouns were (i) lemmatized, (ii) assigned to three (imaginary) customers and a virtual requirements engineer then (iii) merged into one vocabulary set. Each customer interacts with the virtual requirements engineer as depicted in Figure 4.  

The customer initiates the interaction by sending all his nouns to the virtual requirements engineer, who then identifies the mutual occurring nouns and sends those back to the customer. The virtual requirements engineer continues the interaction by sending the customer all of his nouns, after which the customer creates the interaction result. Figure 5. shows the formula, according to which the customer creates the interaction result. A cooperation factor (0..1) determines to what extent the imaginary customer is willing to accept the requirements engineer's nouns to the interaction result. The customer communicates the interaction result nouns to the virtual requirements engineer, who then appends them to his own nouns (knowledge base). After each interaction, the virtual requirements engineer is reset to his initial state. The interaction result nouns as well as the requirements engineer nouns are One Hot Encoded in relation to the vocabulary set. These One Hot Encoded vectors are a mathematical representation of the nouns in this experiment. This Experiment is repeated multiple times for different cooperation factors.

\begin{figure*}[!hb]
    \centering
	\includegraphics[clip,width=0.8\linewidth]{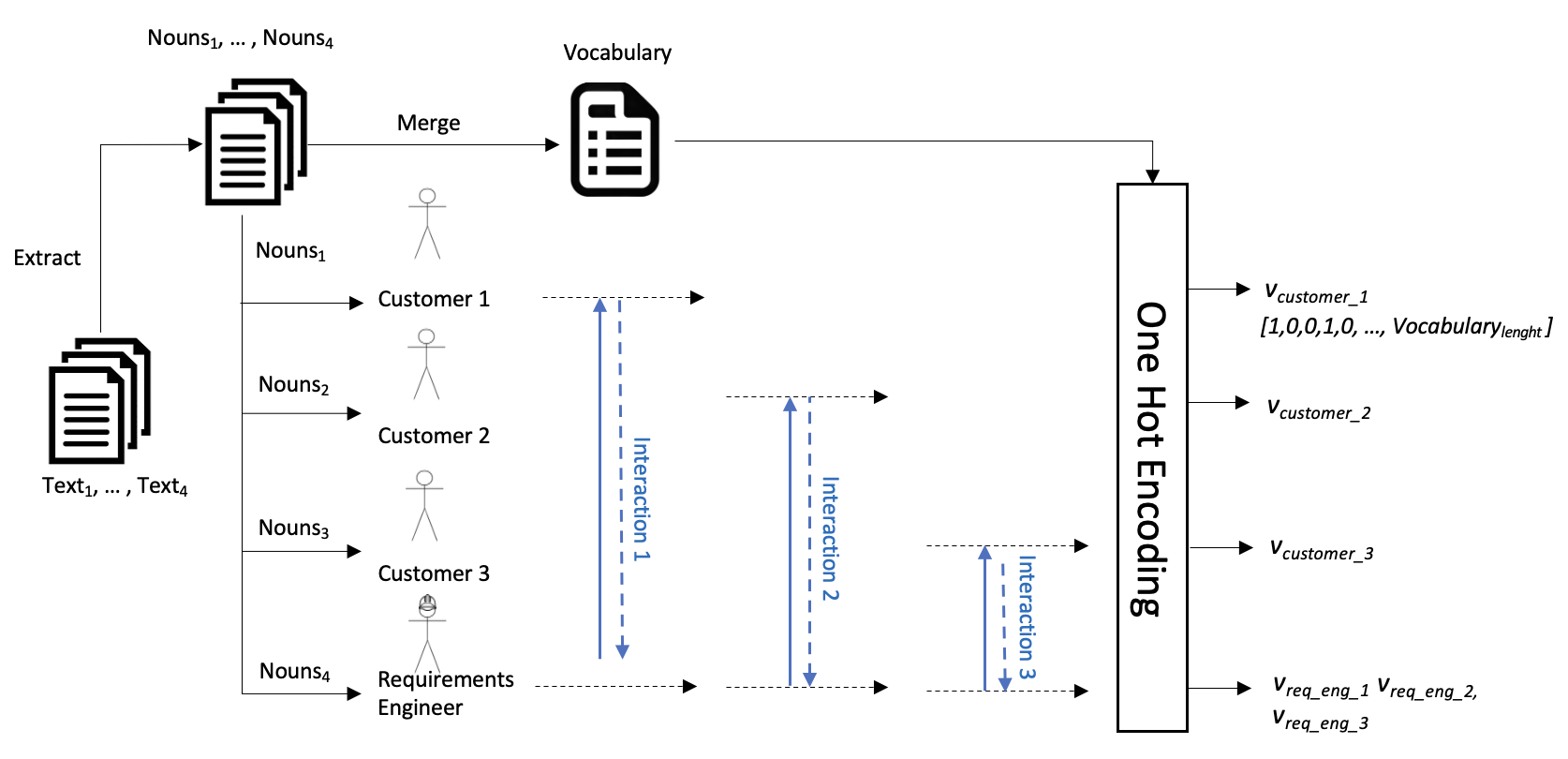}
	\caption{Experiment outline.}
	\label{fig: Figure 3.}       % Give a unique label
\end{figure*}

\hfill \break
\textit{Hypothesis 1:} A higher cooperation factor leads to higher similarity between the interaction results, for all three customers. A cooperation factor of/close to zero would cause heterogeneous results. \textit{Hypothesis 2:} Both, the customer and the virtual requirements engineer will learn from each other throughout the course of the interaction. Their resulting vectors will be more similar to each other after the interaction, then they were before. 

\hfill \break
The corresponding Software Repository for this experiment, can be found at this URL: \textit{https://github.com/VasiliySeibert/hicss23}

\begin{figure}[!htb]
    \centering
	\includegraphics[clip,width=0.5\linewidth]{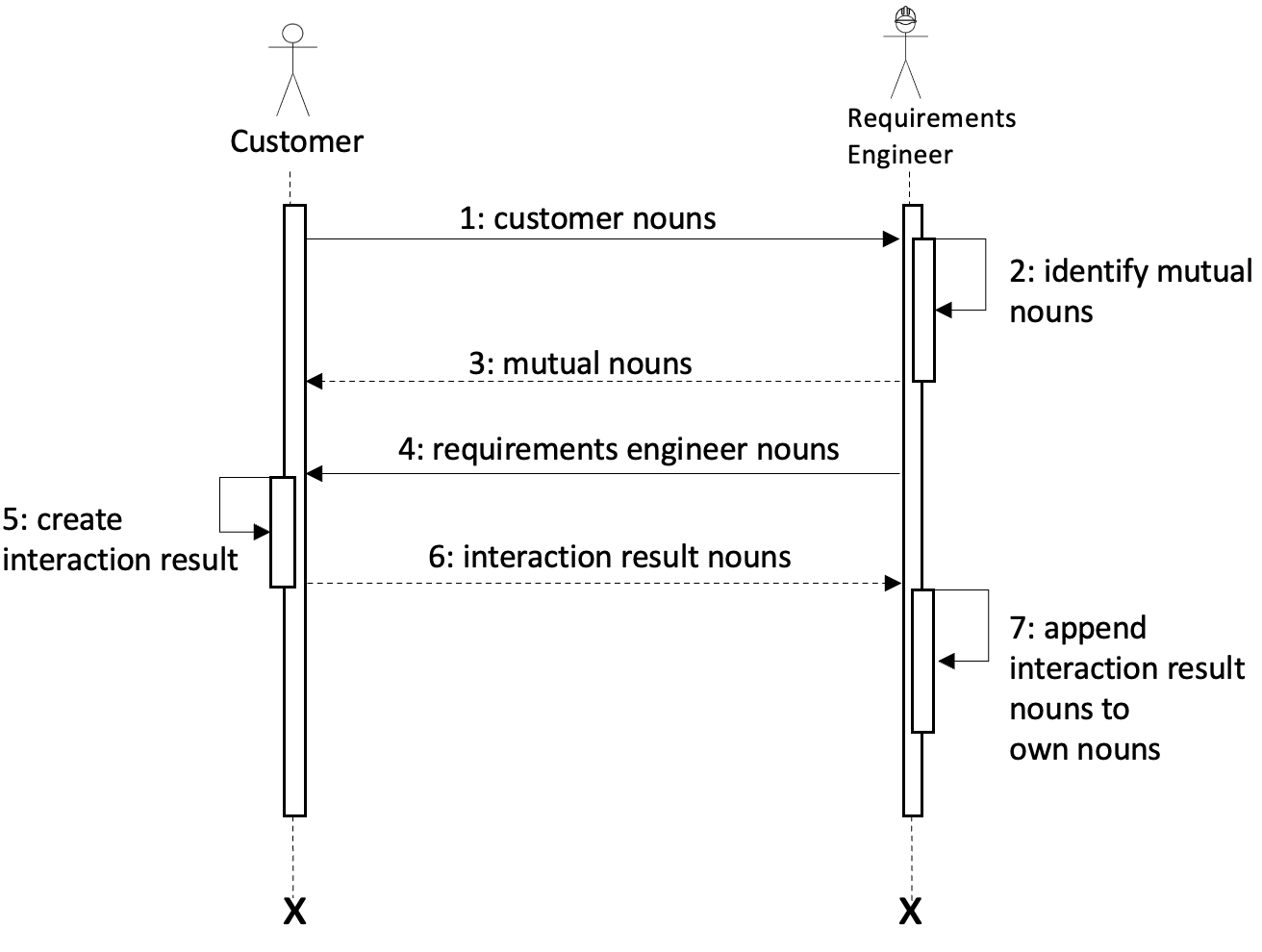}
	\caption{Experiment: Simplistic interaction.}
	\label{fig: Figure 4.}       % Give a unique label
\end{figure}

\begin{figure}[!htb]
    \centering
	\includegraphics[clip,width=0.6\linewidth]{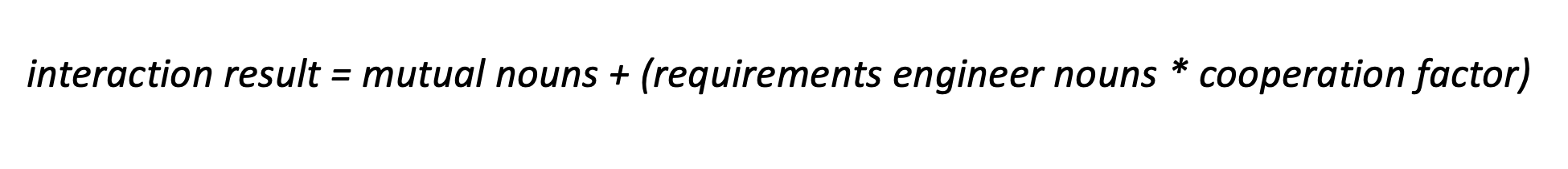}
	\caption{Experiment: creation of the interaction result nouns.}
	\label{fig: Figure 5.}       % Give a unique label
\end{figure}

\section{Results}

By comparing the similarity of the three interaction result nouns (that resulted from the interaction of the customers with the virtual requirements engineer, across different cooperation factors) a higher similarity can be observed over a rising cooperation factor. The similarity of the interaction result nouns is determined by calculating the cosine similarity between the corresponding vectors, which are a mathematical representation of the One Hot Encoded nouns.  A cooperation factor close to 0 causes the interaction result nouns to be heterogeneous. A cooperation factor close to 1 causes the interaction result nouns to be homogeneous. 

The interaction result is created according to the formula depicted in Figure 5. This means that the interaction result contains at least the mutual occurring nouns, even if the cooperation factor is 0 ($ interaction result = mutual nouns$). With a rising cooperation factor, the interaction result is enriched by a fraction of additional requirements engineer nouns. A cooperation factor of 1 would mean that the interaction result contains the mutual nouns as well as all of the additional requirements engineer nouns ($interaction result = mutual nouns \bigcup requirements engineer nouns$).

\begin{figure}[!htb]
    \centering
	\includegraphics[clip,width=0.5\linewidth]{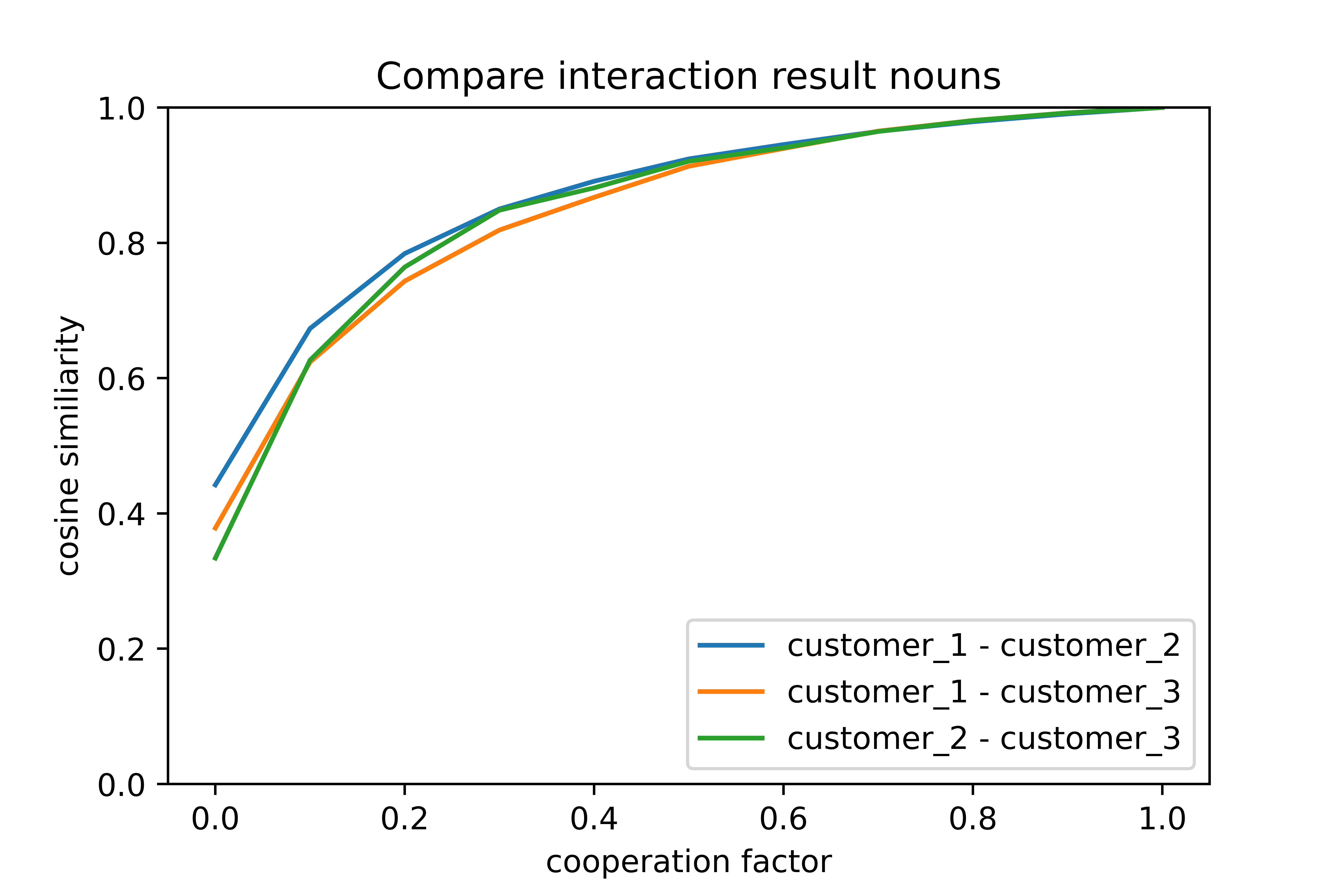}
	\caption{Experiment: Comparing the interaction result nouns.}
	\label{fig: Figure 6.}       % Give a unique label
\end{figure}

\section{Discussion}

\textbf{What does this experiment demonstrate?} The presented experiment is supposed to communicate (a fraction) of the general interaction blueprint idea. It shows how a mutual problem/domain understanding can emerge through (a simplistic) dialogue. It demonstrates the difference between the proposed interaction blueprint and a pure model extraction approach (again, which is the favored approach in today's research). Hypothesis 1 stated that:

\hfill \break
\textit{A higher cooperation factor leads to higher similarity between the interaction results, for all three customers. A cooperation factor of/close to zero would cause heterogeneous results.}
\hfill \break

This polarity maps to real world scenarios, in which (even for individual software solutions) best practices and reoccurring concept structures might be beneficial and can be suggested by an experienced requirements engineer. The degree to which a customer is willing to cooperate determines how much the customer's solution is in line with solutions from comparable scenarios. On the other hand, a low cooperation factor will cause the interaction result to be highly individual and therefore heterogeneous compared with similar scenarios, which should not represent a disadvantage. The cooperation factor is merely an aspect that is yet to be considered in today's research. Hypothesis 2 stated:

\hfill \break
\textit{Both,
the customer and the virtual requirements engineer
will learn from each other throughout the course of the
interaction. Their resulting vectors will be more similar
to each other after the interaction, then they were before.}
\hfill \break

Imagine the virtual requirements engineer being a Large Language Model (LLM), that generates Text by relying on its underlying Word Embedding. The generated text is based on the corpus of texts that it was trained on. It is highly questionable, that a given customer shares the underlying semantic understanding in every instance (e.g. regarding the similarity of Words, \cite{pennington-etal-2014-glove}). For this exact reason, this paper argues that there cannot be a data driven text generation approach that lives up to the task of generating a satisfying domain model from mere descriptions across different customers, fully automatically. That is, because every novelty/abstraction that is generated by the LLM and not double-checked by the customer, poses the risk of causing costly mistakes and misunderstandings. As for the customer, a perfect initial understanding of the problem / domain model would eliminate the need for computer aided software requirements analysis in the first place (or requirements analysis in general for that matter). In real world scenarios, a requirements engineer is hired for his knowledge and expertise (this aspect is also disregarded by current research in this area).

\textbf{What does not this experiment demonstrate?} The proposed interaction blueprint holds potential that is not demonstrated in the experiment. This includes the customer's ability to reflect his initial problem understanding, making adjustments and genuine compromise. The requirements engineer lacks the ability to abstract the discussed concepts and communicate the resulting domain model to the customer. Furthermore, this experiment does not show how the requirements engineer would gather general knowledge throughout every interaction. Since the purpose of this experiment is to demonstrate (a fraction) of the general idea, it does not indicate any potential technologies that could be relevant for its realization.
\hfill \break

\textbf{What influence could this proposed interaction blueprint have on future research?} Ideally, this interaction blueprint motivates a dialogue based, computer aided software requirements analysis. Current advancements in natural language processing and generative AI in general, indicate that this branch of research could experience rapid advancements in a foreseeable future. However, this paper argues that there is a need to move away from a \textit{magical black box expectation} that would extract domain models perfectly. Instead, moving towards a dialogue based approach that recognizes the individuality that is an undeniable part of requirements engineering. Only through dialogue, the results become explainable and can therefore live up to a higher level of trust and acceptance.
\hfill \break

\textbf{What challenges come with the proposed interaction blueprint?} Three obvious challenges for the proposed interaction blueprint are acceptance, evaluation and comparability. This paper must be accepted by a research community to have any effect/influence. It can be seen as a starting point that opens up wide possibilities to contribute and conduct research. Since the proposed interaction blueprint promotes individuality, creativity and compromise defining a learning problem and corresponding metrics for evaluation might be difficult. To extend the previous challenge, providing a standardized dataset that could be shared within the community is not trivial (how could one provide a dataset of a dialogue that could take different directions?).

\section{Conclusion}

This paper reflects on current research that is conducted towards the extraction of models from natural language descriptions. It argues that these approaches have inherent weaknesses for they (a) assume an initial problem understanding that is perfect and (b) they leave no room for customer feedback. Motivated by professional, real-world collaboration settings between requirements engineers and customers, this paper proposes an interaction blueprint that aims for dialogue based, computer aided software requirements analysis. Compared to mere model extraction approaches, this interaction blueprint encourages individuality, creativity and genuine compromise by design. A simplistic Experiment was conducted to showcase (a fraction) of the general idea. Through that experiment, this paper demonstrates how the willingness of the customer to accept novel concepts and let go of details that were considered relevant before can lead to heterogeneous/homogeneous results, even if the underlying problem scenarios are similar. The experiment also demonstrates how both, the customer and the requirements engineer can learn from each other throughout the course of the interaction. With these two results, the simplistic experiment already considered two relevant aspects of requirements engineering that are not considered in today's research. This paper discusses the experiment as well as the proposed interaction blueprint and argues, that advancements in natural language processing and generative AI might lead to significant progress in this branch of research in a foreseeable future. However, for that, there is a need to move away from a magical black box expectation that would extract domain models perfectly. Instead, moving towards a dialogue based approach that recognizes the individuality that is an undeniable part of requirements engineering. Only through dialogue, the results become explainable and can therefore live up to a higher level of trust and acceptance.

\bibliographystyle{unsrtnat}
%\bibliography{references}  %%% Uncomment this line and comment out the ``thebibliography'' section below to use the external .bib file (using bibtex) .


\begin{thebibliography}{6}
\providecommand{\natexlab}[1]{#1}
\providecommand{\url}[1]{\texttt{#1}}
\expandafter\ifx\csname urlstyle\endcsname\relax
  \providecommand{\doi}[1]{doi: #1}\else
  \providecommand{\doi}{doi: \begingroup \urlstyle{rm}\Url}\fi

\bibitem[Ahmed et~al.(2022)Ahmed, Ahmed, and Eisty]{survey_1}
Sharif Ahmed, Arif Ahmed, and Nasir~U Eisty.
\newblock Automatic transformation of natural to unified modeling language: A
  systematic review.
\newblock In \emph{2022 IEEE/ACIS 20th International Conference on Software
  Engineering Research, Management and Applications (SERA)}, pages 112--119.
  IEEE, 2022.

\bibitem[Abdelnabi et~al.(2021)Abdelnabi, Maatuk, and Hagal]{survey_2}
Esra~A Abdelnabi, Abdelsalam~M Maatuk, and Mohammed Hagal.
\newblock Generating uml class diagram from natural language requirements: A
  survey of approaches and techniques.
\newblock In \emph{2021 IEEE 1st International Maghreb Meeting of the
  Conference on Sciences and Techniques of Automatic Control and Computer
  Engineering MI-STA}, pages 288--293. IEEE, 2021.

\bibitem[Abdelnabi et~al.(2020)Abdelnabi, Maatuk, Abdelaziz, and
  Elakeili]{libya}
Esra~A Abdelnabi, Abdelsalam~M Maatuk, Tawfig~M Abdelaziz, and Salwa~M
  Elakeili.
\newblock Generating uml class diagram using nlp techniques and heuristic
  rules.
\newblock In \emph{2020 20th International Conference on Sciences and
  Techniques of Automatic Control and Computer Engineering (STA)}, pages
  277--282. IEEE, 2020.

\bibitem[Deeptimahanti and Babar(2009)]{umgar}
Deva~Kumar Deeptimahanti and Muhammad~Ali Babar.
\newblock An automated tool for generating uml models from natural language
  requirements.
\newblock In \emph{2009 IEEE/ACM International Conference on Automated Software
  Engineering}, pages 680--682. IEEE, 2009.

\bibitem[Elallaoui et~al.(2018)Elallaoui, Nafil, and
  Touahni]{elallaoui2018automatic}
Meryem Elallaoui, Khalid Nafil, and Raja Touahni.
\newblock Automatic transformation of user stories into uml use case diagrams
  using nlp techniques.
\newblock \emph{Procedia computer science}, 130:\penalty0 42--49, 2018.

\bibitem[Pennington et~al.(2014)Pennington, Socher, and
  Manning]{pennington-etal-2014-glove}
Jeffrey Pennington, Richard Socher, and Christopher Manning.
\newblock {G}lo{V}e: Global vectors for word representation.
\newblock In \emph{Proceedings of the 2014 Conference on Empirical Methods in
  Natural Language Processing ({EMNLP})}, pages 1532--1543, Doha, Qatar,
  October 2014. Association for Computational Linguistics.
\newblock \doi{10.3115/v1/D14-1162}.
\newblock URL \url{https://aclanthology.org/D14-1162}.

\end{thebibliography}

\end{document}